\documentstyle[11pt,amssymb,epsf]{article}

\def\ben{\begin{equation}}
\def\een{\end{equation}}

\let\a=\alpha    
    
  \let\n=\nu

\let\C=\Chi

\def\nn{\nonumber} \def\bd{\begin{document}} \def\ed{\end{document}}
\def\ds{\documentstyle} \let\fr=\frac \let\bl=\bigl \let\br=\bigr
\let\Br=\Bigr \let\Bl=\Bigl
\let\bm=\bibitem
\let\na=\nabla
\let\pa=\partial \let\ov=\overline
\newcommand{\be}{\begin{equation}}
\newcommand{\ee}{\end{equation}}
\def\ba{\begin{array}}
\def\ea{\end{array}}
\def\ft#1#2{{\textstyle{{\scriptstyle #1}\over {\scriptstyle #2}}}}
\def\fft#1#2{{#1 \over #2}}
\def\del{\partial}
\def\vp{\varphi}
\def\sst#1{{\scriptscriptstyle #1}}
\def\oneone{\rlap 1\mkern4mu{\rm l}}
\def\td{\tilde}
\def\wtd{\widetilde}
\def\ie{\rm i.e.\ }
\def\dalemb#1#2{{\vbox{\hrule height .#2pt
        \hbox{\vrule width.#2pt height#1pt \kern#1pt
                \vrule width.#2pt}
        \hrule height.#2pt}}}
\def\square{\mathord{\dalemb{6.8}{7}\hbox{\hskip1pt}}}
\newcommand{\ho}[1]{$\, ^{#1}$}
\newcommand{\hoch}[1]{$\, ^{#1}$}
\newcommand{\bea}{\begin{eqnarray}}
\newcommand{\eea}{\end{eqnarray}}
\newcommand{\ra}{\rightarrow}
\newcommand{\lra}{\longrightarrow}
\newcommand{\Lra}{\Leftrightarrow}
\newcommand{\ap}{\alpha^\prime}
\newcommand{\bp}{\tilde \beta^\prime}
\newcommand{\tr}{{\rm tr} }
\newcommand{\Tr}{{\rm Tr} }
\def\0{{\sst{(0)}}}
\def\1{{\sst{(1)}}}
\def\2{{\sst{(2)}}}
\def\3{{\sst{(3)}}}
\def\4{{\sst{(4)}}}
\def\5{{\sst{(5)}}}
\def\6{{\sst{(6)}}}
\def\7{{\sst{(7)}}}
\def\8{{\sst{(8)}}}
\def\n{{\sst{(n)}}}
\def\cA{{{\cal A}}}
\def\cB{{{\cal B}}}
\def\cF{{{\cal F}}}
\def\tV{\widetilde V}
\def\tW{\widetilde W}
\def\tH{\widetilde H}
\def\tE{\widetilde E}
\def\tF{\widetilde F}
\def\tA{\widetilde A}
\def\im{{{\rm i}}}
\def\tY{{{\wtd Y}}}
\def\ep{{\epsilon}}
\def\vep{{\varepsilon}}
\def\R{\rlap{\rm I}\mkern3mu{\rm R}}
\def\bD{{{\bar D}}}

\def\R{\rlap{\rm I}\mkern3mu{\rm R}}
\def\bD{{{\bar D}}}
\def\R{{{\Bbb R}}}
\def\C{{{\Bbb C}}}
\def\H{{{\Bbb H}}}
\def\CP{{{\Bbb C}{\Bbb P}}}
\def\RP{{{\Bbb R}{\Bbb P}}}
\def\Z{{{\Bbb Z}}}
\def\bA{{{\Bbb A}}}
\def\bB{{{\Bbb B}}}
\def\bC{{{\Bbb C}}}
\def\bD{{{\Bbb D}}}
\def\bE{{{\Bbb E}}}
\def\bZ{{{\Bbb Z}}}
\def\Re{{{\frak{Re}}}}
\def\Im{{{\frak{Im}}}}
\def\cosec{{\,\hbox{cosec}\,}}
\def\Gm{{\Gamma_{\!\! -}}}
\def\Gp{{\Gamma_{\!\! +}}}
\def\stan{{standard }}
\def\nonstan{{supernumerary }}

\newcommand{\tamphys}{\it Center for Theoretical Physics,
Texas A\&M University, College Station, TX 77843}

\newcommand{\upenn}{\it Department of Physics and Astronomy,\\ University
of Pennsylvania, Philadelphia, PA 19104}

\newcommand{\brussels}{\it Physique Th\'eorique et Math\'ematique, 
Universit\'e Libre de Bruxelles,\\ Campus Plaine C.P. 231, B-1050
Bruxelles, Belgium} 

\newcommand{\auth}{H. L\"u}

\begin{document}
\begin{flushright}
MCTP-02-22\\
April  2002\\
\hfill{\bf hep-th/0204231}\\
\end{flushright}

\vspace{10pt}

\begin{center}

{\large {\bf New $G_2$ Metric, D6-branes and Lattice Universe}}

\vspace{30pt}
\auth

\vspace{20pt}
{\it Michigan Center for Theoretical Physics\\
University of Michigan, Ann Arbor, Michigan 48109}


\vspace{40pt}

\underline{ABSTRACT}
\end{center}
 
   We construct a new (singular) cohomogeneity-three metric of $G_2$
holonomy.  The solution can be viewed as a triple intersection of
smeared Taub-NUTs.  The metric comprises three non-compact radial-type
coordinates, with the principal orbits being a $T^3$ bundle over $S^1$.
We consider an M-theory vacuum (Minkowski)$_4\times {\cal M}_7$ where
${\cal M}_7$ is the $G_2$ manifold.  Upon reduction on a circle in the
$T^3$, we obtain the intersection of a D6-brane, a Taub-NUT and a
6-brane with R-R 2-form flux.  Reducing the solution instead on the
base space $S^1$, we obtain three intersecting 6-branes all carrying
R-R 2-form flux.  These two configurations can be viewed as a classical
flop in the type IIA string theory.  After reducing on the full
principal orbits and the spatial world-volume, we obtain a
four-dimensional metric describing a lattice universe, in which the
three non-compact coordinates of the $G_2$ manifold are identified
with the spatial coordinates of our universe.

{\vfill\leftline{}\vfill
\vskip 10pt 
\footnoterule {\footnotesize 
Research is supported in full by DOE grant DE-FG02-95ER40899.
}

\pagebreak
\setcounter{page}{1}

\section{Introduction}

       Seven-dimensional manifolds of $G_2$ holonomy have long been
known to exist.  The construction of explicit non-compact $G_2$
metrics began ten years ago, when asymptotically conical metrics of
cohomogeneity one were found \cite{brysal,gibpagpop}.  The physical
interest of $G_2$ manifolds has increasingly significantly with the
discovery of M-theory, because they are the most natural compactifying
spaces from the eleven-dimensional point of view.  It is expected that
M-theory compactified on a $G_2$ manifold gives rise to an ${\cal
N}=1$ super Yang-Mills theory in $D=4$ \cite{acharaya}.  The $G_2$
manifold with principal orbits $S^3\times S^3$ provides a geometrical
demonstration of the classical flop of the type IIA superstring theory
\cite{atmava}.  In \cite{atiwit}, M-theory dynamics on a $G_2$
manifold were discussed.

     Recently, a large class of new metrics of $G_2$ holonomy have
been obtained \cite{cglp6fun}-\cite{ccglpw}, following the
construction of the first examples of asymptotically locally conical
spin(7) manifolds \cite{newspin7}.  These examples have non-abelian
isometry groups.  $G_2$ metrics with nilpotent isometry groups were
also constructed in \cite{gilupost}, which can be obtained by taking
the Heisenberg or Euclidean limits of the non-abelian examples.
Whilst it is of great interest to construct regular $G_2$ metrics,
physically, it is essential to have an appropriate singularity
structure to give rise to chiral fermions in $D=4$ \cite{achwit,csu}.

        In section 2, we construct a new non-compact cohomogeneity
three metric with $G_2$ holonomy.  The metric has three non-compact
radial-type coordinates, with the principal orbits being a $T^3$
bundle over $S^1$.  The isometry group of the metric is a
four-dimensional nilpotent Lie group.  The metric has either power-law
singularities or delta-function singularities.  The solution can be
viewed as the intersection of three smeared Taub-NUTs. When one of the
Taub-NUT charges is set to zero, the metric describes a product of
$S^1$ with a six-dimensional non-compact Calabi-Yau manifold.

       In section 3, we consider an M-theory vacuum
(Minkowski)$_4\times {\cal M}_7$, where ${\cal M}_7$ is the $G_2$
manifold.  We show that by dimensionally reducing the solution on one
of the circles in the $T^3$, we obtain a type IIA configuration with
one D6-brane, one Taub-NUT and one 6-brane with an R-R 2-form flux.
On the other hand, if we reduce the solution on the base space $S^1$,
we obtain an intersection of three 6-branes all carrying R-R 2-form
flux.  These two configurations can be viewed as the classical flop in
type IIA string theory on a non-compact six-dimensional K\"ahler
manifold with a nilpotent isometry group.  The origin of the flop is
that the $T^3$ bundle over $S^1$ principal orbits can also be viewed
as $S^1$ bundle over $T^3$.

       In section 4, we perform a Kaluza-Klein reduction on the full
principal orbits and the spatial world-volume.  We obtain three
perpendicularly intersecting membranes in $D=4$, describing a lattice
universe. In this picture, the three non-compact coordinates of the
$G_2$ manifold are identified with the spatial coordinates of our
universe.  We conclude the letter in section 5.

\section{New $G_2$ metric}
\label{g2metsec}

The metric ansatz is given by
\bea
ds_{7}^2 &=& H_1\, dx_1^2 + H_2\, dx_2^2 + H_3\, dx_3^2 + 
H_1\, H_2\, H_3\, dz_4^2 + H_3\, H_1^{-1}\, 
(dz_1 + H_1'\, z_2\, dz_4)^2\nn\\
&&+ H_1\, H_2^{-1}\, (dz_2 + H_2'\, z_3\, dz_4)^2 +
 H_2\, H_3^{-1}\, (dz_3 + H_3'\, z_1\, dz_4)^2\,,\label{g2metric}
\eea
where $H_1$, $H_2$ and $H_3$ are functions of $x_1$, $x_2$ and $x_3$
respectively.  The prime on $H_i'$ denotes a derivative with respect to
the argument of $H_i$;
\be 
H_1'=\del_{x_1} H_1\,,\qquad H_2'=\del_{x_2} H_2\,,\qquad
H_3'=\del_{x_3} H_3\,.
\ee
The natural vielbein basis is
\bea
&&e^0=\sqrt{H_1\, H_2\, H_3}\, dz_4\,,\qquad
e^1=\sqrt{H_1}\, dx_1\,,\qquad e^2=\sqrt{H_2}\, dx_2\,,\qquad
e^3=\sqrt{H_3}\, dx_3\,,\nn\\
&&e^4=\sqrt{H_3\, H_1^{-1}}\, (dz_1 + H_1'\, z_2\, dz_4)\,,\qquad
e^5=\sqrt{H_1\, H_2^{-1}}\,(dz_2 + H_2'\, z_3\, dz_4)\,,\nn\\
&&e^6=\sqrt{H_2\, H_3^{-1}}\, (dz_1 + H_3'\, z_1\, dz_4)\,.
\label{vielbein}
\eea
The associative 3-form in this basis is given by
\be
\Phi = e^{016} + e^{024} + e^{035} + e^{125} - e^{134} +
e^{236} - e^{456}\,,\label{3form}
\ee
where $e^{ijk}=e^i\wedge e^j\wedge e^k$.  The metric (\ref{g2metric})
has $G_2$ holonomy if and only if $\Phi$ is closed and co-closed.  We
find that the closure and co-closure of $\Phi$ implies that
\be
H_i''=0\,,\qquad i=1,2,3,
\ee
implying that
\be
H_1=1 + m_1\, x_1\,,\qquad
H_2=1 + m_2\, x_2\,,\qquad
H_3=1 + m_3\, x_3\,.\label{sol1}
\ee
Here the constant 1 is included so that $H_i$ does not vanish when
$m_i=0$.  Clearly the metric has a power-law singularity whenever any
of the $H_i$ vanishes.  The metric can also be recast in a
``co-moving'' frame,
\bea
ds_7^2&=& dr_1^2 + dr_2^2 + dr_3^2 +
\ft94(m_1\,m_2\,m_3\,r_1\,r_2\,r_3)^{2/3}\, dz_4^2+ 
\left(\fft{m_3\,r_3}{m_1\,r_1}\right)^{2/3}
(dz_1 + m_1\, z_2\,dz_4)^2\nn\\
&&+\left(\fft{m_1\,r_1}{m_2\,r_2}\right)^{2/3}
(dz_2 + m_2\, z_3\,dz_4)^2 
   +\left(\fft{m_2\,r_2}{m_3\,r_3}\right)^{2/3}
(dz_3 + m_3\, z_1\,dz_4)^2\,.
\label{sol1.5}
\eea
The fibration in the $z_1$, $z_2$ and $z_3$ coordinates implies that
the constants $m_i$ are quantised, namely \cite{fibre}
\be
m_1=n_1 \fft{L_1}{L_2\, L_4}\,,\qquad
m_2=n_2 \fft{L_2}{L_3\, L_4}\,,\qquad
m_3=n_3 \fft{L_3}{L_1\, L_4}\,,
\ee
where $n_i$ are integers and $L_i$ are the periods of the $z_i$.  For
simplicity, we can set $L_i=\ell_p$ where $\ell_p$ is the Plank
length, and then $m_i=n_i/\ell_p$.

      The metric has a power-law singularity when any of the $H_i$
vanishes.  This can be avoided by instead taking
\be
H_i=1 + \sum_\a m_i^\a |x_i - x_i^\a|\,.\label{sol2}
\ee
such that $H_i$ is positive definite.  However, in doing so, we have
introduced delta function singularities.

      When all three of the $H_i'$ are non-vanishing, the metric
describes three intersecting Taub-NUTs with three independent
non-vanishing smeared charges.  The metric has three non-compact
coordinates $x_1$, $x_2$ and $x_3$.  The principal orbits are $T^3$
bundle over $S^1$; they are parameterised by the coordinates $(z_1,
z_2, z_3)$ and $z_4$ respectively.  The metric is of cohomogeneity
three since it depends explicitly on the three non-compact coordinates
$x_i$.  Despite the dependence on the $(z_1,z_2,z_3)$ coordinates,
they, together with $z_4$, parameterise a four-dimensional nilpotent
Lie group ${\cal G}$, which is the isometry group of the metric, and
thus the four-dimensional principal orbits are homogeneous.

      When two of the $H_i'$ vanish, the metric describes a direct
product of Euclidean 3-space and a smeared Taub-NUT.  if instead only
one of $H_i'$ vanishes, in which case the metric was obtained in
\cite{lattice}, it describes a product of an $S^1$ with a Calabi-Yau
6-manifold.  To see this in detail, let us set $H_3=1$.  The metric of
the Calabi-Yau manifold is then given by
\bea
ds_{6}^2 &=& H_1\, dx_1^2 + H_2\, dx_2^2 + H_1\, H_2\,
dz_4^2 +  H_2\, \, dz_3^2\nn\\
&&+H_1^{-1}\, (dz_1 + H_1'\, z_2\, dz_4)^2
+ H_1\, H_2^{-1}\, (dz_2 + H_2'\, z_3\, dz_4)^2\,.
\label{d6metric}
\eea
and the K\"ahler form is given by
\be
J=e^{0}\wedge e^5 - e^1\wedge e^4 + e^2\wedge e^6\,,
\ee
where the vielbein is given by (\ref{vielbein}) with $H_3=1$.

\section{Intersecting D6-branes}
\label{d6sec}

     Having obtained the new $G_2$ metric, one may consider an
M-theory vacuum solution given by the direct product of Minkowski
4-spacetime and the $G_2$ manifold, namely
\be
ds_{11}^2 = - dt^2 + dw_1^2 + dw_2^2 + dw_3^2 + ds_7^2\,.
\ee
The solution can be viewed as a triple intersection of smeared
Taub-NUTs, with the metric represented by the diagram

\bigskip\bigskip
\centerline{
\begin{tabular}{c|ccccccccccc}
&$t$ & $w_1$ & $w_2$ & $w_3$ & $x_1$ & $x_2$ & $x_3$ &
$z_1$ & $z_2$ & $z_3$ & $z_4$ \\ \hline
$H_1(x_1)$ &$\times$ & $\times$ & $\times$ & $\times$ & $-$ &
$\times$ & $\times$ & $*$ & $-$ & $\times$ & $-$ \\
$H_2(x_2)$&$\times$ & $\times$ & $\times$ & $\times$ & $\times$ 
& $-$ &$\times$ & $\times$ & $*$ & $-$ & $-$ \\
$H_3(x_3)$&$\times$ & $\times$ & $\times$ & $\times$ & $\times$ & 
$\times$ &$-$ & $-$ & $\times$ & $*$ & $-$ \\
\end{tabular}}
\bigskip

\centerline{Diagram 1. Triple intersections of Taub-NUTs. Here
$\times$, $-$ and $*$ denote}
\centerline{the world-volume, transverse space, and
fibre coordinates respectively.\phantom{X}} 
\bigskip

There is a $U(1)$ isometry for each of the $z_i$ coordinates, and so
we can reduce the metric on any $z_i$ to obtain a solution in type IIA
theory.  The $z_i$ for $i=1,2,3$ are equivalent, and hence the
reduction can be discussed using $z_1$ as a representative.  The
resulting type IIA solution is given by
\bea
e^{\ft16\phi}\,ds_{10}^2 &=& -dt^2 + dw_1^2 + dw_2^2 + dw_3^2 +
H_1\, dx_1^2 + H_2\, dx_2^2 + H_3\,dx_3^2  + H_1\, H_2\,
H_3\,dz_4^2\nn\\
&& + H_1\, H_2^{-1}\, (dz_2 + H_2'\, z_3\, dz_4)^2 +
H_2\,H_3^{-1}\, dz_3^2 + H_3\, H_1\, (H_1'\, z_2)^2\, dz_4^2\nn\\
&&-W^{-1}\,(H_3\, H_1^{-1}\, H_1'\, z_2\, dz_4 -
H_2\, H_3^{-1}\, H_3'\, z_4\, dz_3)^2\,,\nn\\
e^{\phi}&=&W^{\ft34}\,,\qquad
W=H_3\, H_1^{-1} + H_2\, H_3^{-1}\, (H_3'\, z_4)^2\,,\nn\\
A_\1 &=& W^{-1}\, (H_3\, H_1^{-1}\, H_1'\, z_2\, dz_4 -
H_2\, H_3^{-1}\, H_3'\,z_4\, dz_3)\,.
\eea
Note that before performing the Kaluza-Klein reduction, we have made a
coordinate transformation $z_3\rightarrow z_3 - H_3'\, z_1\, z_4$ in
the metric (\ref{g2metric}).  Clearly, the solution describes an
intersection of three objects.  The one parameterised by $H_1$ is a
smeared D6-brane, and the one parameterised by $H_2$ is a Taub-NUT.
The one associated with $H_3$ is a 6-brane carrying an R-R 2-form
flux, but it differs from a standard D6-brane.

      We can instead reduce the solution on the $z_4$ coordinate,
giving the type IIA solution
\bea
&&e^{\ft16\phi}\,ds_{10}^2 = -dt^2 + dw_1^2 + dw_2^2 + dw_3^2 +
H_1\, dx_1^2 + H_2\, dx_2^2 + H_3\, dx_3^2\nn\\
&& \qquad\qquad\quad+ H_3\, H_1^{-1}\, dz_1^2 + H_1\, H_2^{-1}\, dz_2^2 +
H_2\, H_3^{-1}\, dz_3^2\nn\\
&&\qquad\qquad\quad - W^{-1}\,(H_3\, H_1^{-1}\, H_1'\, z_2\, dz_1 +
H_1\, H_2^{-1}\, H_2'\, z_3\, dz_2 +
H_2\, H_3^{-1}\, H_3'\, z_1\, dz_3)^2\,,\nn\\
&&e^{\phi}=W^{\ft34}\,,\quad
W=H_1\, H_2\, H_3 + H_3\, H_1^{-1}\, (H_1'\, z_2)^2 +
H_1\, H_2^{-1}\, (H_2'\, z_3)^2 +
H_2\, H_3^{-1}\, (H_3'\, z_1)^2\,,\nn\\
&&A_\1=W^{-1}\,(H_3\, H_1^{-1}\, H_1'\, z_2\, dz_1 +
H_1\, H_2^{-1}\, H_2'\, z_3\, dz_2 +
H_2\, H_3^{-1}\, H_3'\, z_1\, dz_3)\,.
\eea
The solution describes three intersecting 6-branes all carrying
R-R 2-form flux.  These 6-branes are different from the usual
D6-brane coming from the reduction of the fibre coordinate of a
Taub-NUT in $D=11$.

       The two configurations arising from the reduction on $z_1$ or
$z_4$ can be viewed as a classical flop in the type IIA string theory
on the non-compact K\"ahler manifold.  The flop in $D=10$ can be
geometrically explained by the fact that the $T^3$ bundle over $S^1$
principal orbits of the four-dimensional nilpotent Lie group ${\cal
G}$ can also be described as an $S^1$ bundle over $T^3$.  However, the
two descriptions are somewhat different.  In the latter case, the
fibre is the circle group in ${\cal G}$ generated by $\ft{\del}{\del
z_4}$.  In the former case, the fibre is not the orbit of a
three-dimensional subgroup of ${\cal G}$ because $\ft{\del}{\del z_i}$
for $i=1,2,3$ are not themselves Killing vectors; we must add a
multiple of $\ft{\del}{\del z_4}$.  In fact the flop involves
interchanging the fibre and base spaces of the $U(1)$ fibration.  This
is analogous to the flop discussed in \cite{atmava}.

\section{Lattice universe}
\label{latticesec}

      The new $G_2$ metric (\ref{g2metric}) that we have obtained is
in fact inspired by the four-dimensional intersecting membrane
solution that describes the lattice universe
\cite{lattice}.\footnote{In \cite{triple}, a triply quasi-periodic
Gibbons-Hawking metric was obtained.}  There has been experimental
evidence suggesting that the network of galaxy superclusters and voids
seems to form a three-dimensional lattice with a spacing of about $120
h^{-1}\,\,Mpc$ (where $h^{-1}$ is the Hubble constant in units of $100
km\,\, s^{-1}\,\, Mpc^{-1}$) \cite{lattice1,lattice2,lattice3}.  In
\cite{lattice}, an M-theory solution was constructed to describe such
a lattice structure, which can be realised by considering non-standard
brane intersections of two M5-branes and one Taub-NUT, or two
Taub-NUTs and one M5-brane.  In the latter case, turning off the
M5-brane charge causes the solution to reduce to the product of
5-dimensional Minkowski spacetime and the non-compact Calabi-Yau
manifold given in (\ref{d6metric}).
  
        In section 2, we obtained the new $G_2$ metric (\ref{g2metric})
by adding an extra fibration on the seventh coordinate.  This
procedure follows the general prescription of obtaining $G_2$
manifolds from six-dimensional K\"ahler manifolds, described in
detail in \cite{almost}.

        If we reduce the M-theory solution on the world-volume spatial
coordinates $w_i$ and also on the $T^3$ bundle over $S^1$ principal
orbits, we obtain three perpendicularly intersecting membranes in
$D=4$, with the metric
\be
ds_{4}^2 = (H_1\, H_2\, H_3)^{\ft12}\, (-dt^2 + H_1\, dx_1^2 +
H_2\, dx_2^2 + H_3\, dx_3^2)\,.\label{lattice}
\ee
This metric was first obtained in \cite{lattice}, although it was
supported by very different field strength.  The functions of $H_i$ in
this case are given by (\ref{sol2}) describing periodic arrays of
intersecting membranes.

      In this static cosmological model the spatial world-volume and
the $T^3$ bundle over $S^1$ principal orbits are viewed as an internal
space, whilst the three non-compact coordinates $x_1$, $x_2$ and $x_3$
of the $G_2$ manifold are identified with the spatial coordinates of
our universe.  This is rather natural since the principal orbits are
clearly compact, and the spatial world-volume can be wrapped on a
compact space such as $T^3$.  Although it is not likely that the
metric (\ref{lattice}) describes our actual universe, since it
preserves ${\cal N}=1$ supersymmetry; it is nevertheless rather
suggestive that the lattice structure should emerge from a metric with
$G_2$ holonomy.

\section{Conclusions}

        In this letter, we constructed a new cohomogeneity-three
metric with $G_2$ holonomy.  It has three radial-type coordinates,
with the principle orbits being a $T^3$ bundle over $S^1$.  The
solution can be viewed as three intersecting Taub-NUTs.  We performed
Kaluza-Klein reduction on the $S^1$ and instead on a circle in the
$T^3$.  The two resulting type IIA configurations can be viewed as a
classical flop in type IIA string theory on a non-compact K\"ahler
six-manifold.  Although the type IIA solutions do not describe the
triply intersecting D6-branes advocated in \cite{csu,wit} for the
realisation of chiral fermions, it is nevertheless of interest to
investigate further if chirality could arise from the singularities of
our $G_2$ metric.  The metric provides a concrete example for studying
such an issue, since it can be viewed as the lifting of a D6-brane
configurations in $D=10$.

       If we perform a Kaluza-Klein reduction on the entire
four-dimensional full principal orbits and the spatial world-volume,
we obtain triply intersecting membranes in $D=4$, describing a lattice
universe.  The construction takes full advantage of the non-compact
nature of the manifold, in that the three non-compact coordinates are
precisely identified with the spatial coordinates of our universe.  It
is of great interest to investigate further the significance of such a
configuration arising from a $G_2$ manifold.

\section*{Acknowledgement}

    We are grateful to Mirjam Cveti\v c for extensive discussions on
triply intersecting D6-branes and chiral fermions, to Gary Gibbons for
pointing out the isometry group of the metric, and to Chris Pope and
Justin V\'azquez-Poritz for discussions.

\end{document}